\documentclass[letter,longauth]{aa}

\usepackage{lscape}
\usepackage{graphicx}
\usepackage{lineno}

\newcommand{\kms}{km\,s$^{-1}$}
\newcommand{\masyr}{mas\,yr$^{-1}$}

\newcommand{\msun}{M$_{\odot}$}

\newcommand{\vrot}{$v_\mathrm{e} \sin i$}

\newcommand{\s}{$\sigma$}

\newcommand{\gaia}{{\em Gaia}}

\def\aap{A\&A}

\def\apjs{ApJS}
\def\apjl{ApJ}

\begin{document} 
   \title{The VLT-FLAMES Tarantula Survey}

   \subtitle{Observational evidence for two distinct populations\\ of massive runaway stars in 30 Doradus}
   \titlerunning{Runaway stars in 30~Doradus}

   \author{H. Sana
          \inst{1}
          \and
O.~H.~Ram\'irez-Agudelo\inst{2} \and 
V.~H\'enault-Brunet\inst{3} \and  
L.~Mahy\inst{1,20} \and 
L.~A.~Almeida\inst{4,21} \and 
A.~de~Koter\inst{5,1} \and 
J.~M.~Bestenlehner\inst{6} \and  
C.~J.~Evans\inst{7} \and   
N.~Langer\inst{8} \and 
F.~R.~N.~Schneider\inst{9,10} \and  
P.~A.~Crowther\inst{6} \and 
S.~E.~de~Mink\inst{11,5} \and 
A.~Herrero\inst{12} \and 
D.~J.~Lennon\inst{12} \and 
M.~Gieles\inst{13,14} \and 
J.~Ma\'{ii}z~Apell\'aniz\inst{15} \and 
M.~Renzo\inst{16} \and 
E.~Sabbi\inst{17} \and 
J.~Th.~van~Loon\inst{18} \and 
J.~S.~Vink\inst{19}  
          }

   \institute{Institute of Astronomy, KU Leuven, Celestijnlaan 200D, 3001 Leuven, Belgium\\
              \email{hugues.sana@kuleuven.be}
         \and 
German Aerospace Center (DLR), Institute for the Protection of Terrestrial Infrastructures, Rathausallee 12, 53757 Sankt Augustin, Germany 
         \and
Department of Astronomy and Physics, Saint Mary's University, 923 Robie Street, Halifax, NS B3H 3C3, Canada
            \and 
Escola de Ci\^encias e Tecnologia, Universidade Federal do Rio Grande do Norte, Natal - RN, 59072-970, Brazil
\and 
Anton Pannenkoek Astronomical Institute, University of Amsterdam, 1090 GE Amsterdam, The Netherlands
\and 
Department of Physics \& Astronomy, Hounsfield Road, University of Sheffield, Sheffield, S3 7RH, UK
\and 
European Space Agency (ESA), ESA Office, Space Telescope Science Institute, 3700 San Martin Drive, Baltimore, MD 21218, USA 
\and 
Argelander-Institut für Astronomie, der Universit{\"a}t Bonn, Auf dem Hügel 71, 53121 Bonn, Germany
	\and 
Heidelberger Institut f\"{u}r Theoretische Studien, Schloss-Wolfsbrunnenweg 35, 69118, Heldelberg, Germany
 	\and
Astronomisches Rechen-Institut, Zentrum f{\"u}r Astronomie der Universit{\"a}t Heidelberg, M{\"o}nchhofstr.\ 12-14, 69120 Heidelberg, Germany
\and  
Max Planck Institut f\"{u}r Astrophysik, Karl-Schwarschild-Strasse 1, 85748 Garching, Germany
\and 
Instituto de Astrof\'{i}sica de Canarias, 
           C/ V\'{i}a L\'{a}ctea s/n, E-38200 La Laguna, Tenerife,
           Spain
\and
ICREA, Pg. Llu\'{i}s Companys 23, E08010 Barcelona, Spain
\and Institut de Ci\`{e}ncies del Cosmos (ICCUB), Universitat de Barcelona (IEEC-UB), Mart\'{i} Franqu\`{e}s 1, E08028 Barcelona, Spain
\and 
Centro de Astrobiolog\'{i}a (-INTA). Campus ESAC. Camino bajo del castillo s/n. 28\,692 Villanueva de la Ca\~nada, Madrid, Spain
\and
Center for Computational Astrophysics, Flatiron Institute, New York, NY 10010, USA
Department of Physics, Columbia University, New York, NY 10027, USA 
\and 
Space Telescope Science Institute, 3700 San Martin Drive, Baltimore, MD 21218, United States
\and
Lennard-Jones Laboratories, Keele University, ST5 5BG, UK
\and 
Armagh Observatory, College Hill, Armagh, BT61 9DG, Northern Ireland, UK
\and 
Royal Observatory of Belgium, Av.\ Circulaire 3, 1180 Uccle, Belgium
\and 
Programa de P\'os-gradua\c{c}\~ao em F\'isica, Universidade do Estado do Rio Grande do Norte, Mossor\'o - RN, 59610-210, Brazil
             }
 \date{Received; accepted}
 \abstract
 {The origin of massive runaway stars is an important unsolved problem in astrophysics. Two main scenarios have been proposed -- dynamical ejection or release from a binary at the first core collapse -- but their relative contribution remains heavily debated.}
 {Taking advantage of two large spectroscopic campaigns towards massive stars  in 30~Doradus, we aim to provide observational constraints on the properties of the O-type runaway population in the most massive active star-forming region in the Local group. }
 {We use radial velocity measurements of the O-type star populations in 30 Doradus obtained by  the VLT-FLAMES Tarantula Survey and the Tarantula Massive Binary Monitoring to identify single and binary O-type runaways. We further discuss the rotational properties of the detected runaways and qualitatively compare observations with expectations of ejection scenarios.}
 {We identify 23  single and one binary O-type runaway objects, most of them located outside the main star-forming regions in 30~Doradus. We find an overabundance of rapid rotators ($v_\mathrm{e} \sin i > 200$~\kms) among the runaway population, hence providing an explanation of the observed overabundance of rapidly rotating stars in the 30~Doradus field. Considerations of the projected rotation rates and runaway line-of-sight velocities reveal a conspicuous absence of rapidly rotating ($v_\mathrm{e} \sin i > 210$~\kms), fast moving ($v_\mathrm{los} > 60$~\kms) runaway stars in our sample, and strongly suggest the presence of two different populations of runaway stars: a population of rapidly-spinning but slowly moving runaway stars and a population of fast moving but slowly rotating ones. These are detected with a ratio close to 2:1 in our sample.}
 {
We argue that slowly moving but rapidly spinning runaway stars result from binary ejections, while rapidly moving but slowly spinning runaways could result from dynamical ejections. Given that detection biases will more strongly impact the slow-moving runaway population, our results suggest that the binary evolution
scenario dominates the current massive runaway star population in 30 Doradus. 
} 


   \maketitle
%
\section{Introduction}

Massive runaway stars \citep{Bla61} hurtle through space with speeds of tens to a couple of hundreds of \kms. During their lifetimes they can travel large distances (up to 1000s of pc) from their birthplaces, meaning that they can inject radiative and mechanical energy and processed chemical elements deep into the interstellar -- sometimes intergalactic -- medium \citep{CeK09, AAR20}. 

Two distinct mechanisms have been proposed to explain their origin: dynamical interaction in a dense star cluster \citep{Bla61, GiB86}, or release from a close binary system when the companion explodes as a supernova \citep{Zwi57, Sto91}. Both of these are believed to occur in nature \citep{HdBdZ00}, but their relative efficiency, the resulting ejection rates and velocity distributions  have been a topic of debate over the past 50 years \citep{Bla61, HdBdZ01, JOD10, JOP20}. Understanding these mechanisms and their relative contributions is pivotal to quantify the feedback of massive stars in the broader context of reionization and the chemical enrichment of galaxies \citep{CoK12}.

\begin{figure}
\centering
\includegraphics[width=\columnwidth]{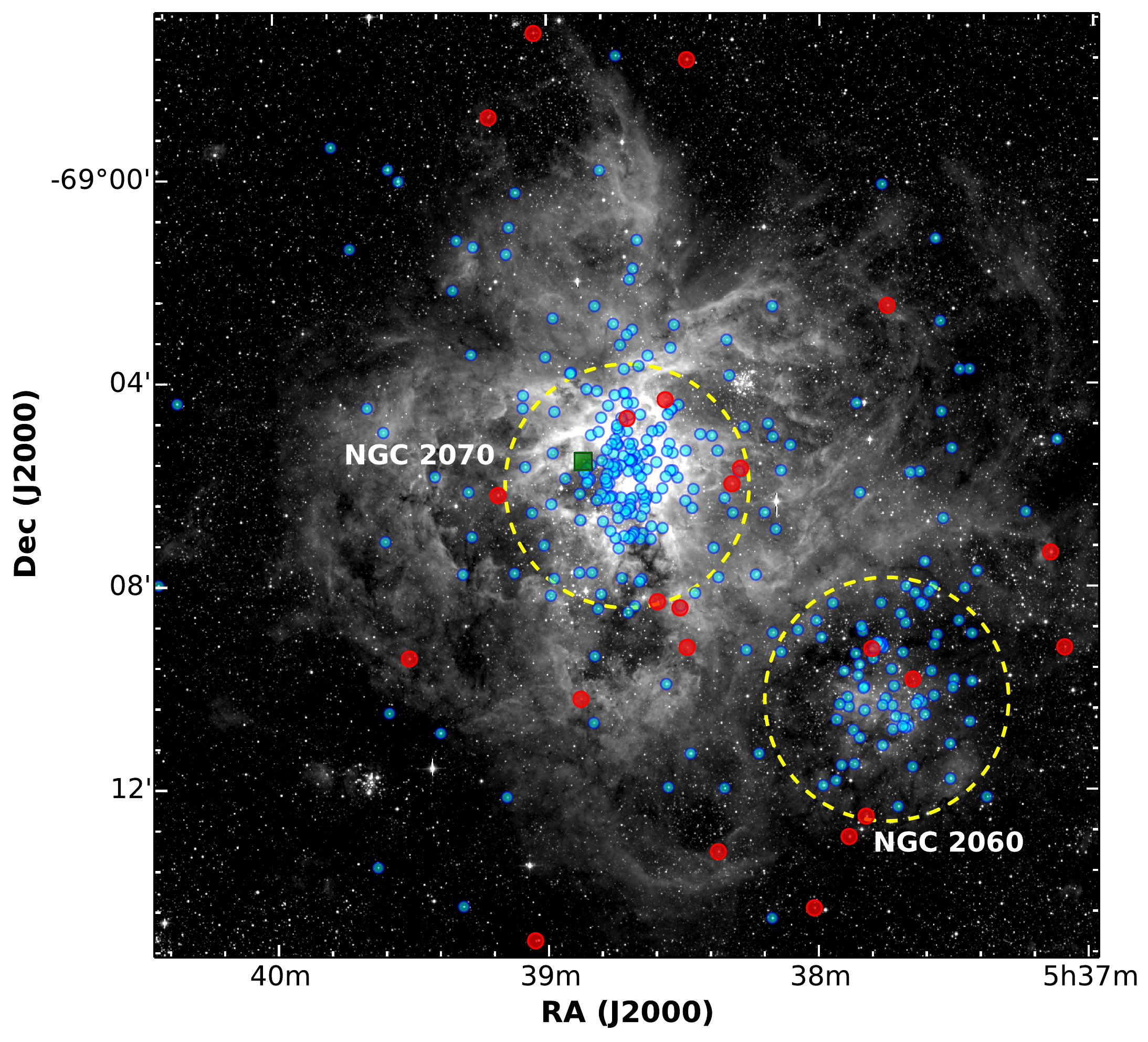}
\caption{$V$-band Wide Field Imager (WFI) mosaic of the 30~Doradus region.  O-type stars runaways single (red) and binary (green) identified in this work are spread throughout the field of view, while the non-runaway O-type stars (blue circles) cluster within the NGC~2060 and NGC~2070 OB associations (dashed 2.4'-radius circles, equiv.\ 36~pc in radius).}
\label{fig: fov}
\end{figure}

The 30~Doradus region (a.k.a.\ 30 Dor) in the Large Magellanic Cloud (LMC) is the nearest strong starburst and contains one of the richest massive-star populations in the Local Group of galaxies. 30~Dor therefore gives us a unique laboratory to obtain statistically meaningful observational constraints on the properties of massive stars, including those of runaways. In particular, the massive young cluster R136 (at the centre of 30~Dor) is expected to provide a suitably dense environment for dynamical interactions to occur, but is likely too young to have produce any supernova. Moreover, the metallicity of the LMC is $\sim$50\% solar \citep[e.g.,][]{BAK12, LCM19}, so 30~Dor provides a valuable window into the properties of massive stars in conditions comparable to those during the peak of star formation in the Universe.

The short lifetimes of O-type stars allow us to obtain a (close to) real-time snapshot of the production of massive runaways in 30~Dor, while also minimizing possible interlopers that would originate from outside our studied region. Here we employ measurements from multi-epoch spectroscopy of 339 massive O-type stars obtained at the Very Large Telescope (VLT) in the framework of the VLT-FLAMES Tarantula Survey \citep[VFTS;][]{ETHB11}, combined with follow-up spectroscopy obtained over 20 months to monitor 74 of the 118 identified O-type binaries \citep{AST17}.  Finally, we contrast the specific line-of-sight velocities ($v_\mathrm{los}$) of the identified runaways and their projected spins in an attempt to separate out their production channels.


\begin{table*}
\centering
\caption{Systemic velocities and 1\s\ radial-velocity dispersions from the two analysis methods (see text) and runaway (RW) fractions of the single O-star populations in 30~Dor.}
\label{tab: cluster}
\begin{tabular}{lcccc}
\hline
\hline
Population & Sample & Sample Stat.  & Gaussian fit & RW fraction\\
           & size   & (\kms) & (\kms) & (\%)\\
\hline
Overall   & 185  & 271.7 $\pm$            12.6 & 270.7 $\pm$            10.6 & $\phantom{1}$7 $\pm$ 1\\
NGC~2070  &  90  & 267.0 $\pm$ $\phantom{1}$9.6 & 268.2 $\pm$ $\phantom{1}$8.6 & $\phantom{1}$6 $\pm$ 2\\
NGC~2060  &  33  & 277.5 $\pm$ $\phantom{1}$6.3 & 277.3 $\pm$ $\phantom{1}$8.0 & $\phantom{1}$4 $\pm$ 2\\  
Remaining &  62  & 273.2 $\pm$            19.0 & 272.5 $\pm$            14.8 & 23 $\pm$ 5\\  \hline
\end{tabular}
\end{table*}

\section{Observational data} \label{sect: data}

\subsection{The VLT-Flames Tarantula Survey}

The bulk of the data used in this work were acquired as part of the VFTS \citep{ETHB11}, a 160h Large Programme at the European Southern Observatory (ESO). Executed from 2009 to 2010, the VFTS obtained multi-epoch optical spectroscopy of 800 OB-type stars in the 30~Dor region. The targets spanned a field 20$'$ in diameter ($\approx 300$~pc) centred on the R136 cluster (see Fig.~\ref{fig: fov}), but avoids most of the inner 1'-region, hence most of R136 itself, due to crowding \citep[though see ][for an attempt to get closer]{HBES12}. The data were obtained with the Fibre Large Array Multi-Element Spectrograph \citep[FLAMES,][]{PAB02}, using its Medusa fibres to feed the Giraffe spectrograph with light from up to 132 targets simultaneously.
The LR02 and LR03 settings of the low-resolution grating were used to provide
continuous coverage of 3960-5050~\AA\ at a spectral resolving power $\lambda / \Delta \lambda$ of $\sim$7500. The H$\alpha$ region was also observed for each target at $\lambda / \Delta \lambda=15\,000$ with the HR15N setting. 

The observations and data reduction were described by \citet{ETHB11}. Spectral classification, extinction properties, absolute luminosities, absolute radial velocities (RVs), projected rotational velocities ($v_\mathrm{e} \sin i$) and atmospheric parameters of the O-type stars in the survey have been presented in various papers in the VFTS series \citep{SdKdM13, RASDS13, MAEB14, SSSDH14, WSSD14}. Evolutionary parameters (age: $\tau$; current mass: $M_\mathrm{act}$) were estimated for each object using a Bayesian method that matches selected observational constraints to stellar evolutionary models \citep{BdMC11} while accounting for the observational uncertainties \citep{SSE18}. 

O-type spectroscopic binaries and binary candidates were identified from RV variations \citep{SdKdM13}. Of the 339 O-type stars observed by the VFTS, 185 showed no statistically-significant RV variations and are presumed to be single stars (though some of them will inevitably be undetected binaries; see additional considerations below on the binary detection completeness). A further 36 stars displayed significant RV variations, but with peak-to-peak amplitudes of less than 20~\kms. These are either spectroscopic binaries or stars displaying photospheric activity \citep{AST17, SSBC20}. The remaining 118 objects, with peak-to-peak RV variations $>$\,20~\kms, are  considered strong spectroscopic binary candidates \citep{SdKdM13}.

\subsection{The Tarantula Massive Binary Monitoring}

Long-term spectroscopic monitoring of 76 of the 118 O-type binary candidates was obtained over October 2012 to March 2014 in the Tarantula Massive Binary Monitoring programme \citep[TMBM,][]{AST17}. This enabled detailed characterisation of a
representative fraction (65\%) of the known massive spectroscopic binaries in 30~Dor. The TMBM obtained 32 spectra of each system with the LR02 setting, with the observations sampling timescales of days and weeks up to 20 months.

The data were reduced and RVs were estimated using identical methods to those employed for the VFTS data \citep{ETHB11, SdKdM13}. We measured the orbital periods through a Fourier analysis and adjusted the RV curves to obtain the orbital parameters \citep{Cum04}, including a refined orbital period, eccentricity, semi-amplitude of the RV curve, systemic velocity and ephemeris. Further characterisation of the double- and single-line spectroscopic binaries where reported in \citet{MAS20,MSA20} and \citet{SSM2022a,SSM2022b}, respectively.

We rejected four systems with tentative orbital periods $>$510 days (the length of the observational campaign) as strong aliasing did not allow us to firmly derive the orbital properties. The resulting estimates of systemic velocities of the remaining 72 binaries allowed us to search for runaway binary systems using the same approach as for single stars.

\begin{figure}
\centering
\includegraphics[width=\columnwidth]{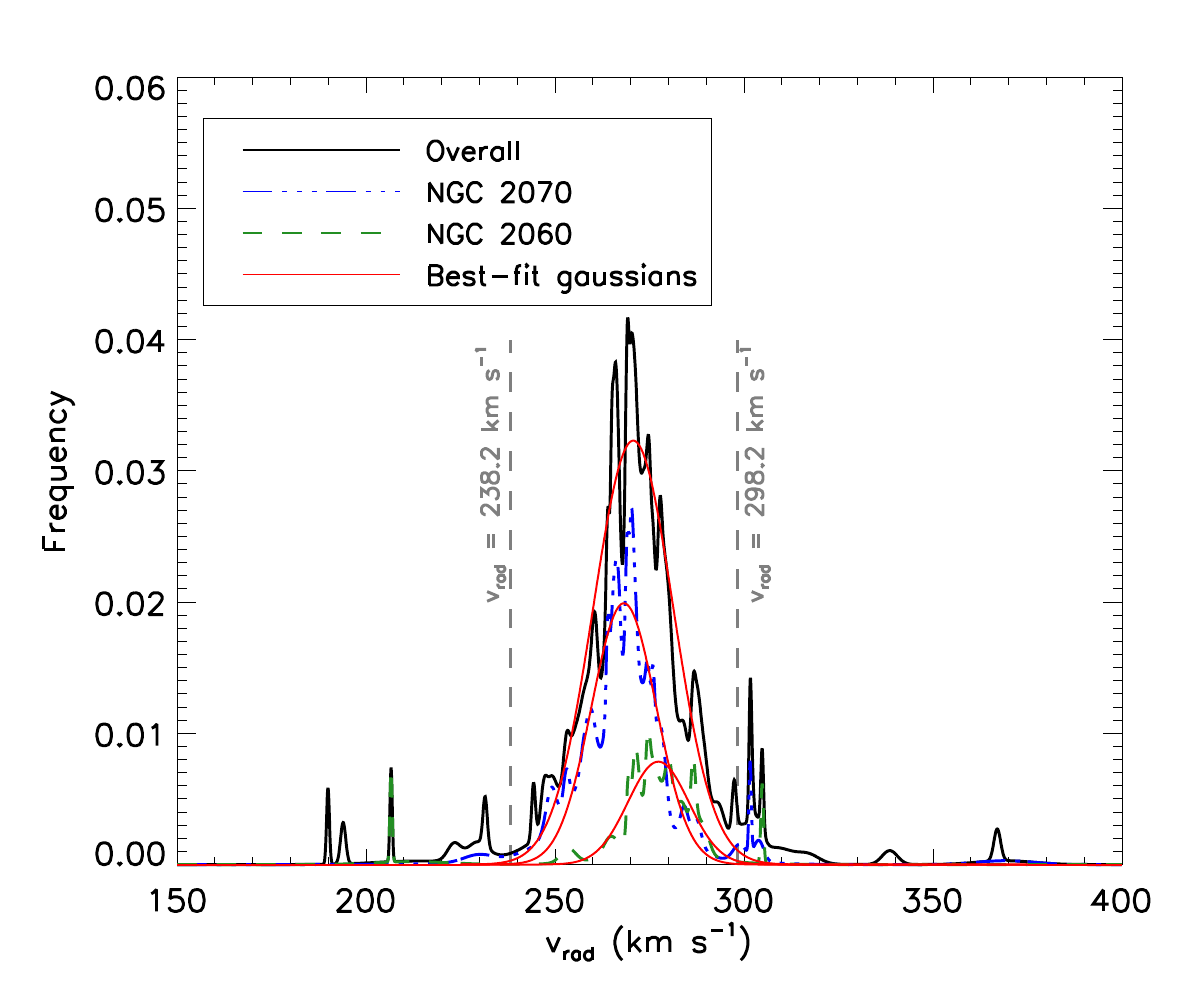}
\caption{Generalized line-of-sight velocity histograms of the O-type stars in 30~Dor together with the best-fit Gaussian distribution function (Table~\ref{tab: cluster}).}
\label{fig: RVhisto}
\end{figure}

\begin{table*}
\centering
\caption{Properties of the O-type runaway stars detected in 30~Dor. Quoted uncertainties are 1\s\ errorbars for the velocities \citep{SdKdM13} and 95\%\ confidence interval for the present-day evolutionary masses ($M_\mathrm{act}$) and ages ($\tau$) \citep{SSE18}, but for VFTS~661 for which the mass estimates is taken from \citet{MAS20}.}
\label{tab: RW}
\begin{tabular}{llllllll}
\hline \hline
 VFTS    &Loca- & $v_\mathrm{los}$    & $\delta v_\mathrm{los}$ & \vrot & Spectral   & $M_\mathrm{act}$ & $\tau$   \\
ID       &tion  & (\kms)              & (\kms)  & (\kms)           &    Classification   &  (\msun) & (Myr) \\
\hline
012       & Out.  & 316.8  $\pm$   3.0  & 48.6 &  306  $\pm$ 31   & O9 IIIn           &  $18.8^{+1.1}_{-0.8}$   &   $5.3^{+0.4}_{-0.5}$   \\ 
016       & Out.  & 189.9  $\pm$   0.4  & 78.3 &   $\phantom{1}$94  $\pm$ 30   & O2 III-If*        &  $91.6^{+11.5}_{-10.5}$   &  $0.7^{+0.1}_{-0.1}$ \\
102       & 2060  & 212.0  $\pm$   9.9  & 56.2 &  610  $\pm$ 61   & O9: Vnnne+        &  $37.2^{+4.6}_{-4.5}$\footnote{Not passing QC of Schneider}   &   $<0.9$\footnote{Not passing QC of Schneider} \\
138$^a$   & Out.  & 297.7  $\pm$   3.4  & 29.5 &  350  $\pm$ 35   & O9 Vn             &  $19.4^{+1.2}_{-1.1}$   &   $1.3^{+1.2}_{-1.3}$ \\
165       & 2060  & 206.6  $\pm$   0.3  & 61.6 &   $\phantom{1}$75  $\pm$ 14   & O9.7 Iab          &  $31.4^{+2.5}_{-3.6}$   &   $4.6^{+0.4}_{-0.4}$ \\
168       & 2060  & 304.8  $\pm$   0.4  & 36.6 &   $\phantom{1}$39  $\pm$ 20   & O8.5 Vz           &  $23.4^{+1.3}_{-1.0}$   &   $3.1^{+0.4}_{-0.7}$ \\
190       & Out.  & 223.2  $\pm$   1.7  & 45.0 &  444  $\pm$ 44   & O7 Vnn((f))p      &  $24.4^{+1.3}_{-2.6}$\footnote{Not passing QC of Schneider}     &   $5.8^{+1.3}_{-1.5}$\footnote{Not passing QC of Schneider}   \\
226       & Out.  & 193.9  $\pm$   0.7  & 74.3 &   $\phantom{1}$64  $\pm$ 10   & O9.7 III          &  $15.6^{+0.6}_{-0.6}$   &   $2.7^{+1.2}_{-1.45}$ \\
285\footnote{Also detected in the proper motion study of \citep{PLvdM18}.} & 2070  & 229.6  $\pm$   3.8  & 38.6 &  622  $\pm$ 62   & O7.5 Vnnn         &  $20.0^{+2.2}_{-1.4}$   &   $1.9^{+2.5}_{-1.8}$ \\
306       & 2070  & 301.6  $\pm$   0.3  & 33.4 &   $\phantom{1}$71  $\pm$ 10   & O8.5 II((f))      &  $30.4^{+3.3}_{-2.2}$   &   $4.4^{+0.4}_{-0.4}$ \\
328       & Out.  & 310.3  $\pm$   3.2  & 42.1 &  238  $\pm$ 10   & O9.5 III(n)       &  $17.2^{+0.8}_{-0.8}$   &  $1.0^{+1.0}_{-1.0}$\\
355       & Out.  & 301.9  $\pm$   0.4  & 33.7 &  134  $\pm$ 26   & O4 V((n))((fc))z  &  $45.8^{+6.7}_{-5.3}$   &  $1.9^{+0.1}_{-0.1}$ \\
356       & Out.  & 338.5  $\pm$   2.0  & 70.3 &  211  $\pm$ 42   & O6: V(n)z         &  $28.4^{+3.1}_{-2.9}$   &  $2.6^{+0.6}_{-0.9}$\\ 
370       & Out.  & 231.5  $\pm$   0.6  & 36.7 &   $\phantom{1}$68  $\pm$ 10   & O9.7 III          &  $16.4^{+0.7}_{-0.8}$   &  $4.3^{+1.0}_{-1.3}$ \\
406       & 2070  & 303.9  $\pm$   1.3  & 35.7 &  356  $\pm$ 36   & O6 nn             &  $32.0^{+5.3}_{-1.4}$   &  $3.3^{+1.0}_{-0.4}$\footnote{Not passing QC of Schneider}  \\
418$^a$   & 2070  & 298.8  $\pm$   0.4  & 30.6 &  133  $\pm$ 13   & O5 V((n))((fc))z  &  $35.0^{+4.7}_{-3.9}$   &  $1.3^{+0.6}_{-0.8}$ \\ 
529$^a$   & 2070  & 295.6  $\pm$  10.8  & 27.4 &  218  $\pm$ 30   & O9.5(n)           &  $17.4^{+1.4}_{-1.2}$   &  $6.4^{+1.2}_{-1.3}$\footnote{Not passing QC of Schneider}  \\
663       & Out.  & 305.4  $\pm$   3.6  & 37.2 &  $\phantom{1}$90  $\pm$ 10   & O8.5 V            &  $21.0^{+1.9}_{-1.7}$   &  $2.3^{+1.3}_{-1.6}$ \\ 
722       & Out.  & 229.4  $\pm$   2.4  & 38.8 &  404  $\pm$ 40   & O7 Vnnz           &  $23.0^{+1.5}_{-1.8}$   &   $3.4^{+1.6}_{-1.2}$\\
724$^a$   & Out.  & 296.8  $\pm$  10.2  & 28.6 &  369  $\pm$ 30   & O7 Vnnz           &  $24.8^{+5.0}_{-5.1}$   &   $3.0^{+1.1}_{-2.5}$ \\
755       & Out.  & 301.9  $\pm$   1.6  & 33.7 &  286  $\pm$ 29   & O3 Vn((f*))       &  $50.8^{+5.4}_{-7.9}$   &   $1.7^{+0.5}_{-0.7}$ \\
761       & Out.  & 367.0  $\pm$   0.9  & 98.8 &  111  $\pm$ 10   & O6.5 V((n))((f))z &  $28.0^{+1.6}_{-1.5}$   &   $1.2^{+0.6}_{-0.8}$ \\
797$^a$   & Out.  & 297.4  $\pm$   0.6  & 29.2 &  138  $\pm$ 20   & O3.5 V((n))((fc)) &  $48.6^{+7.2}_{-5.4}$   &   $1.8^{+0.3}_{-0.3}$ \\
\hline
661$^a$   & 2070  & 294.6  $\pm$   7.1  & 26.4 & \ldots            & O6.5 V(n) + O9.7: V:& 26+17 & \ldots\\
\hline
\end{tabular}
\vspace*{2mm}
\flushleft {\sc notes:} $a$. At 3\s\ from the mean RV value of the complete and NGC~2070 samples but not for NGC~2060 (but also not spatially associated with NGC~2060). 
\end{table*}

\section{Identifying runaway stars through their peculiar line-of-sight velocity} \label{sect: RWdata}

We identify runaway stars as those with estimated RVs that deviate by more than 3$\sigma$ from the systemic RV of the O-star population in the survey. Our analysis proceeds as follows. From the 339 O-type objects in the VFTS \citep{WSSD14}, we select those without significant RV variations, i.e. the 185 presumed single stars. We first measure the systemic and 1\s\ dispersion of the single stars using the unweighted mean and dispersion of the RV sample and a $\kappa-\sigma$ clipping to reject outliers. We also compute generalized RV histograms for the different O-star populations in the observed field, that we fit using Gaussian distributions (Fig.~\ref{fig: RVhisto}). The systemic velocities ($v_\mathrm{sys}$) and dispersions ($\sigma_\mathrm{v}$) obtained from these two methods and for the different O-star populations are in excellent agreement (Table~\ref{tab: cluster}). In doing so, we use the same definition of the NGC\,2070 and NGC\,2060 regions -- the two main O-star associations in the 30~dor region --  as adopted in the VFTS \citep{SdKdM13}.  These are displayed in Fig.~\ref{fig: fov}. The population of  O star outside the two regions are indicated as {\it 'Remaining'} in Table~\ref{tab: cluster}.  In the following, we use the estimate of $\sigma_\mathrm{v}$ obtained from the generalized histograms as they take into account individual error bars.

We flag objects as runaway stars if their line-of-sight velocity ($v_\mathrm{los}$) differs by more than 3\s\ ($25.8$~\kms) from the systemic velocity of NGC~2070 ($v_\mathrm{sys}^{2070}$), which dominates the O-star population of 30~Dor (Fig.~\ref{fig: fov}).  We note that the mean systemic velocities of stars in the NGC~2060 and NGC~2070 differ by $\sim$10~\kms, but that all the runaways identified below are either outside NGC~2060 or have RVs that differ by more than 3\s\ from the systemic velocity of NGC~2060. There is thus no confusion on their runaway nature, even if NGC~2060 is their parent cluster.

The peculiar line-of-sight velocity ($\delta v_\mathrm{los}$) of the runaway stars (both the single stars and binary systems from the TMBM) is thus given by
\begin{equation}
\delta v_\mathrm{los}=|v_\mathrm{los} - v_\mathrm{sys}^{2070}| > v_\mathrm{thres}.
\end{equation}

In total, 23 (presumed) single stars and one binary meet our RV threshold ($v_\mathrm{thres}= 25.8$~\kms), yielding an observed runaway fraction of $7\pm1$\%\ among the O-type population of 30~Dor, where binomial statistics has been used to compute the observational uncertainties. Adopting the canonical criterion \citep{HdBdZ00,HdBdZ01} of a peculiar velocity larger than $30$~\kms\ would reduce the number of single runaway stars by two and remove the only qualifying runaway binary system. However, it would not change the conclusions of this letter. No other runaway stars were identified in the sample with significant RV variations below 20~\kms. 

The spectroscopic binary system that qualifies as a candidate runaway system
is VFTS~661 ($\delta v_\mathrm{los} = 26.4\pm7.1$~\kms). VFTS~661 is a short period ($P\sim 1.3$~d) detached double-line spectroscopic and eclipsing binary. Unfortunately, the large uncertainty on its systemic velocity combined with its proximity to the RV threshold precludes a firm identification. 

Most of the identified runaways have peculiar velocities of up to 60~\kms, but five have even larger velocities of up to 98~\kms\ (Table~\ref{tab: RW}).  The identified runaways are mostly located outside the two main OB associations in the region, NGC~2060 and NGC~2070 (Fig.~\ref{fig: fov}). The fraction of runaway stars in these regions is $4\pm2$\%\ and $6\pm2$\%\ respectively, while it reaches $23\pm5$\%\ in the field outside the two associations. This finding is in line with the runaway nature of the objects and strongly suggests that the massive-star field population in 30~Dor is largely composed of stars ejected from the main associations, albeit some of them with a low velocity and/or with unfavourable line-of-sight motions.

The true number of runaway stars in 30~Dor is indeed expected to be much larger as three main observational effects impact our detection: the fact that our selection method is only sensitive  to runaway stars that have sufficient line-of-sight velocities ($v_\mathrm{los}>25.8$~\kms),  the finite size of our field of investigation (20\arcmin\ in diameter), and the completeness fraction of the survey ($\sim 0.7$).  Contamination of our single star sample by undetected binaries would result in false detections, but this latter bias is largely outweighed by the former three effects. Detailed modelling of these biases is beyond the scope of this letter but a factor of about two to four is to be expected between the detected and true numbers of runaways in the 30~Dor region. 

\begin{figure}
\centering
\includegraphics[width=\columnwidth]{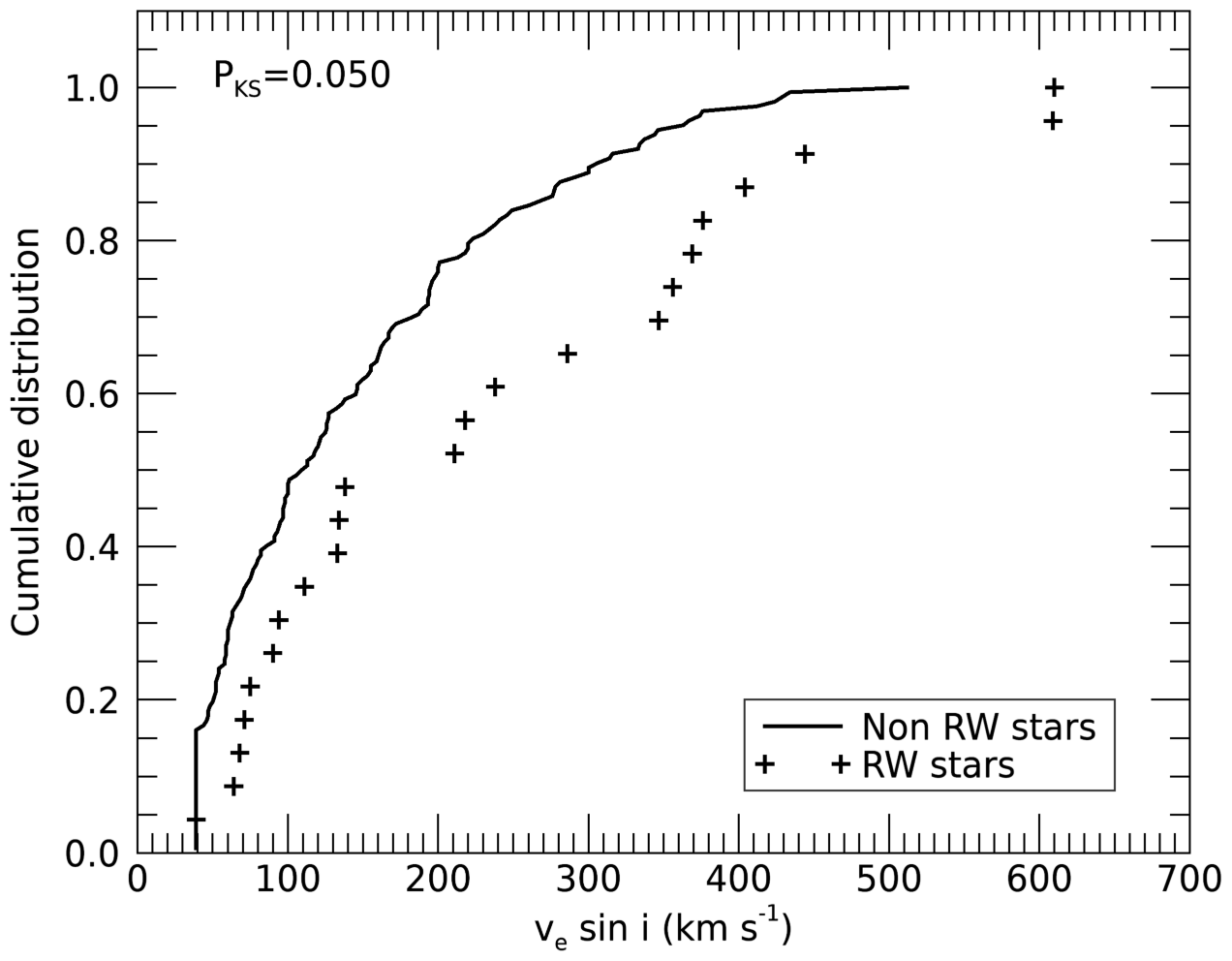}
\caption{Cumulative distribution of the projected rotational velocities (\vrot) of the runaway stars ($+$) compared to that of the non-runaway stars (solid curve). $P_\mathrm{KS}$ indicate the Kolmogorov-Smirnov probability that the two samples are drawn from the same parent distribution.}
\label{fig: cdf_vrot}
\end{figure}

\section{Rotational properties of runaway stars} \label{sect: rot}
\subsection{Projected rotational velocity distribution} \label{sect: rot_obs}
We adopt the projected rotational velocities measured in \citet{RASDS13} for presumably single O stars.  Typical uncertainties are of the order of 20 to 30~\kms. Figure~\ref{fig: cdf_vrot} shows the  cumulative projected spin distributions for stars identified as
runaways and those that are not.
A Kolmogorov-Smirnov test indicates that the two samples show differences with a significance level better than 5\%. The probability of drawing by chance, from the non-runaway sample, a number of fast rotators ($v_\mathrm{e} \sin i > v_\mathrm{cutoff}$) as large as the one observed in the runaway sample drops well below 1\%\ as soon as $v_\mathrm{cutoff}$ is larger than 200~\kms. This again suggests that the two distributions differ significantly. The confidence of the results is consistently better than 2.5\s\ for $v_\mathrm{cutoff}>200$~\kms\ and even reaches 3\s\ above 300~\kms. Thus, there appears to be a cutoff value above which the rotational properties of the runaway and non-runaway populations are incompatible with one another, with the runaway population being preferentially populated by rapidly spinning stars. The overabundance of fast rotators increases exponentially towards larger spinning rates. It is about a factor of two above 200~\kms, a factor of three above 300 and a factor of six above 400~\kms.

Interestingly, rapid rotators are more abundant in the single O-star sample outside the NGC~2060 and 2070 OB associations than within these associations. Indeed, $32\pm6$\%\ among the former have $v_\mathrm{e} \sin i > 200$~\kms\ with only $20\pm3$\%\ among the latter. It seems therefore plausible that this overabundance of fast rotators  outside the main OB associations results from runaway stars spreading across the 30~Dor field-of-view.


\subsection{Rapidly spinning vs. fastly moving runaways} \label{sect: RWdesert}

Simultaneous considerations of the projected rotational ($v_\mathrm{e} \sin i$) and peculiar line-of-sight  ($\delta v_\mathrm{los}$) velocities reveal distinct runaway populations (Fig.~\ref{fig: RWdesert}): a rapidly spinning slow-moving group and a slowly spinning fast-moving group. The absence of rapidly spinning fast-moving runaway stars is significant to better than 99\%\ independent of various assumptions that can be made on the parent runaway velocity distributions. This reveals a zone of avoidance in the  $v_\mathrm{e} \sin i$ -- $\delta v_\mathrm{los}$ parameter space: a runaway desert. The existence of this runaway desert is supported by the {\em Gaia} DR3 proper motion measurements of our sample stars as discussed in the Appendix.

\section{Discussion and conclusions} \label{sect: rot_orig}

\begin{figure}
\centering
\includegraphics[trim= 10 0 40 40, clip,width=\columnwidth]{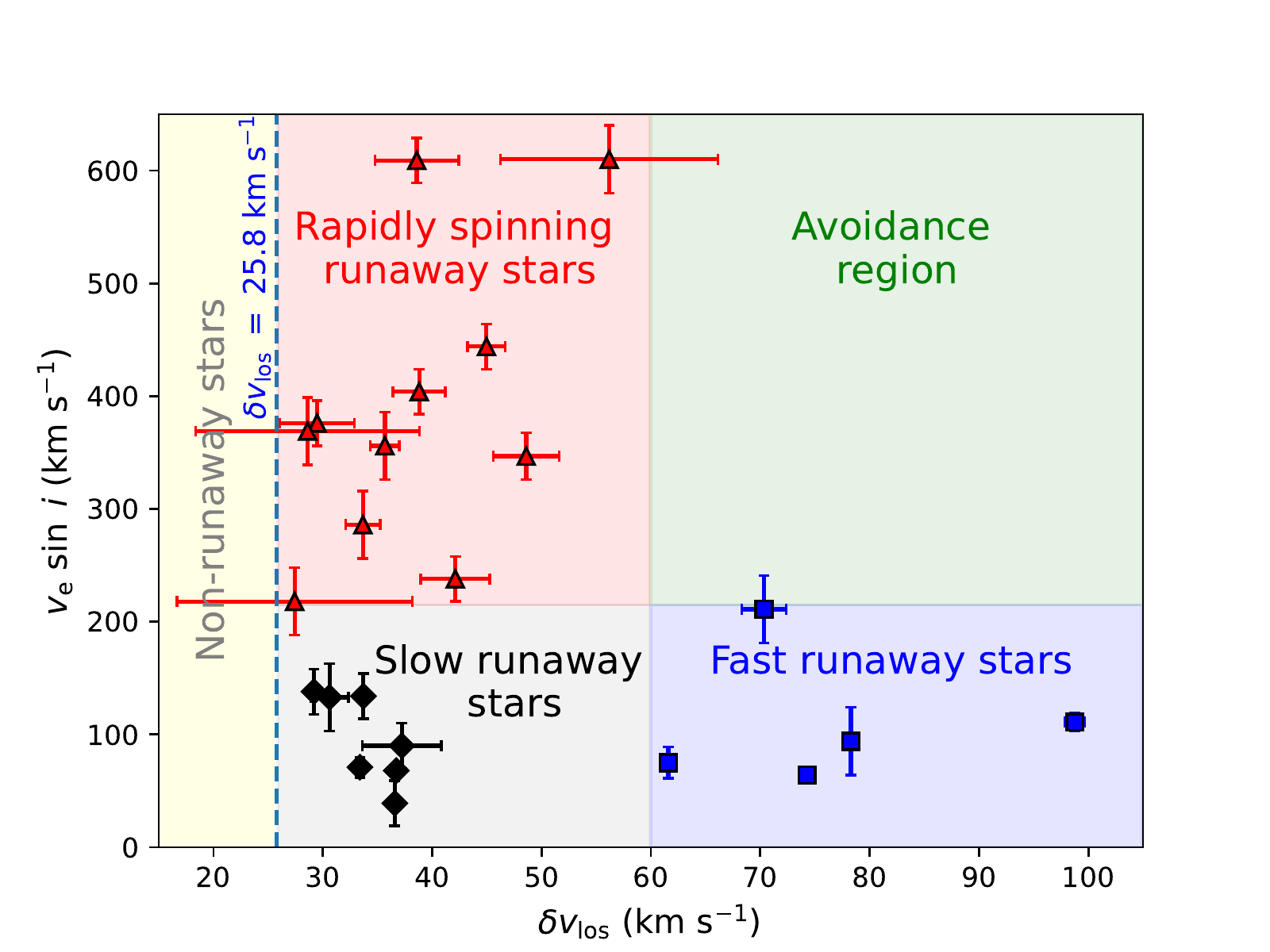}
\caption{Diagram of the projected rotational velocities ($v_\mathrm{e} \sin i$) versus peculiar line-of-sight velocities ($\delta v_\mathrm{los}$) for the runaway stars identified in the 30 Doradus region. }
\label{fig: RWdesert}
\end{figure}
Fast rotation is one of the clearest signatures of post-binary interaction systems \citep{RZdM19}. Under the assumption of continuous star formation, theory predicts \citep{dMLI13} that about 19\%\ of the stars might be spun up to $v_\mathrm{e} \sin i$ $>$ 200~\kms\ either through coalescence (5\%) or mass-transfer (14\%). A small fraction ($\approx 4$\%) of objects are interacting but not significantly spun up (slow case-A interaction Algol-like systems). Aside from merger products \citep{SOP19, SOP20}, some post-interaction products may also show modest or moderate rotation, because of inclination effects or a lack of significant accretion in non-conservative mass-transfer cases. Tidal locking in relatively tight binaries may also prevent the secondary to be significantly spun up \citep{SLM2022}. 
 The latter systems are also among those that have the largest orbital velocity so that a fraction of the high-space velocity runaways may show no significant rotational velocity increase, in qualitative agreement with the distribution of runaway stars in the $v_\mathrm{e} \sin i$ vs.\ $\delta v_\mathrm{los}$ plane (Fig.~\ref{fig: RWdesert}).  


 However, the efficiency of producing fast runaways from binary evolution is model dependant \citep{ELT11,RZdM19,ERR20,SLM2022}. 
Dynamical interactions occur more often in regions of high stellar density \citep{FPZ11}, pointing to R136 at the core of 30~Dor as the most likely origin. The possible ongoing merger between R136 and Sabbi~1 \citep{SLG12} suggests R136 was built from sub-clusters with short timescales for dynamical ejection of O-stars (proportional to the relaxation timescale), possibly creating a favourable environment to further contribute to the dynamically-ejected runaway population. In either case and given its young age \citep{BCCN20, BdKB22}, it is unlikely that R136 has been ejecting stars for more than 2.5 Myr.

With 55\%\ of the runaway O-star sample displaying $v_\mathrm{e} \sin i$ $>$ 200~\kms\ (including the two fastest rotating stars of the entire VFTS  sample), the 30~Dor runaway population presents a statistically-significant overabundance of rapidly-rotating stars compared to its non-runaway population (25\%\ with $v_\mathrm{e} \sin i > 200$~\kms). In addition, the ejection rate of the binary channel increases with age, and the one of the dynamical channel decreases. In regards of the young age of the sample and the fact that detection biases impact comparatively more slow-moving runaways than fast-moving runaways (see Sana et al.\ in prep), our results support a predominance of the binary channel in producing the current and future runaway population in 30~Dor. These results contrast with the recent conclusions of \citet{JOP20} on the origin of runaways in the Small Magellanic Cloud. 

Three of our 11 rapidly spinning -- likely post binary interaction -- runaway stars have masses in the range of 30 to 50 solar masses (Table~\ref{tab: RW}). Ignoring dynamical effects, these high masses suggest that at least some dying  stars sufficiently massive to form black holes must receive natal kicks  that release their companions as runaways. Combined with the recent discovery of X-ray quiet O+BH binaries with significantly different eccentricities \citep{MSS22, SSM2022a, SSM2022b}, this would suggest a diversity of collapse scenarios at play.  Binary-single and binary-binary exchange interactions in dense clusters can also lead to the ejection of a rapidly rotating stars, reducing the need to invoke BH kicks \citep{CT12}.


Finally, the overall limited contribution of dynamical interaction may result from an absence of regions with extremely high stellar density within 30~Dor.  This could suggest that R136 only underwent cluster core collapse recently (or not yet), implying a moderate central density at birth. If confirmed, this would  challenge the efficiency of massive star formation theories that rely on high stellar densities to significantly contribute to the  formation of the most massive stars.

While the {\em Gaia} DR3 measurements support our results, the current precision does not allow us to trace back most of the runaways to their ejection locus \citep{PLvdM18}, but for some rare cases \citep{EWC10,LEvdM18}. Future {\em Gaia} data releases will hopefully provide a sufficient view of stellar motions in  the  three-dimensional space and provide further observational constraints on their space velocity distribution. Comparison with the observational properties of the Milky Way runaway population that is 
being revealed by {\em Gaia}  \citep{MAPGB18} will provide the tools needed to investigate the impact of the environment (metallicity, parent cluster properties) on the properties of massive-star runaways. 

\begin{acknowledgements}
This paper is based on data collected at the European Southern Observatory under program ID 182.D-0222, 090.D-0323 and 092.D-0136. The authors are grateful to  P. Dufton, M. Kennedy, J. Puls, S.~J.~Smartt, N. Walborn and to the VFTS consortium for stimulating discussions. The research leading to these results has received funding from the European Research Council (ERC) under the European Union's Horizon 2020 research and innovation programme (grant agreement numbers 772225: MULTIPLES and 945806: TEL-STARS). M.A. acknowledges support from the Spanish Government Ministerio de Ciencia through grant and PGC2018-095\,049-B-C22. V.H.B. acknowledges the support of the Natural Sciences and Engineering Research Council of Canada (NSERC) through grant RGPIN-2020-05990. M.G. acknowledges support from the Ministry of Science and Innovation (EUR2020-112157, PID2021-125485NB-C22) and from Grant CEX2019-000918-M funded by MCIN/AEI/10.13039/501100011033. L.A.A. thanks the Conselho Nacional de Desenvolvimento Cient\'ifico e Tecnol\'ogico (CNPq) for support through process number 315502/2021-5. FRNS is supported by the Deutsche Forschungsgemeinschaft (DFG, German Research Foundation) under Germany’s Excellence Strategy EXC 2181/1-390900948 (the Heidelberg STRUCTURES Excellence Cluster). This work has made use of data from the European Space Agency (ESA) mission Gaia (https://www.cosmos.esa.int/gaia), processed by the Gaia Data Processing and Analysis Consortium (DPAC, https://www.cosmos.esa.int/web/gaia/dpac/consortium). Funding for the DPAC has been provided by national institutions, in particular the institutions participating in the Gaia Multilateral Agreement. 
\end{acknowledgements}

\bibliography{literature}

\begin{thebibliography}{50}
\expandafter\ifx\csname natexlab\endcsname\relax\def\natexlab#1{#1}\fi

\bibitem[{{Almeida} {et~al.}(2017){Almeida}, {Sana}, {Taylor}, {Barb{\'a}},
  {Bonanos}, {Crowther}, {Damineli}, {de Koter}, {de Mink}, {Evans}, {Gieles},
  {Grin}, {H{\'e}nault-Brunet}, {Langer}, {Lennon}, {Lockwood}, {Ma{\'{\i}}z
  Apell{\'a}niz}, {Moffat}, {Neijssel}, {Norman}, {Ram{\'{\i}}rez-Agudelo},
  {Richardson}, {Schootemeijer}, {Shenar}, {Soszy{\'n}ski}, {Tramper}, \&
  {Vink}}]{AST17}
{Almeida}, L.~A., {Sana}, H., {Taylor}, W., {et~al.} 2017, \aap, 598, A84

\bibitem[{{Andersson} {et~al.}(2020){Andersson}, {Agertz}, \& {Renaud}}]{AAR20}
{Andersson}, E.~P., {Agertz}, O., \& {Renaud}, F. 2020, \mnras, 494, 3328

\bibitem[{{Banerjee} {et~al.}(2012){Banerjee}, {Kroupa}, \& {Oh}}]{BAK12}
{Banerjee}, S., {Kroupa}, P., \& {Oh}, S. 2012, \apj, 746, 15

\bibitem[{{Bestenlehner} {et~al.}(2020){Bestenlehner}, {Crowther},
  {Caballero-Nieves}, {Schneider}, {Sim{\'o}n-D{\'\i}az}, {Brands}, {de Koter},
  {Gr{\"a}fener}, {Herrero}, {Langer}, {Lennon}, {Ma{\'\i}z Apell{\'a}niz},
  {Puls}, \& {Vink}}]{BCCN20}
{Bestenlehner}, J.~M., {Crowther}, P.~A., {Caballero-Nieves}, S.~M., {et~al.}
  2020, \mnras, 499, 1918

\bibitem[{{Blaauw}(1961)}]{Bla61}
{Blaauw}, A. 1961, \bain, 15, 265

\bibitem[{{Brands} {et~al.}(2022){Brands}, {de Koter}, {Bestenlehner},
  {Crowther}, {Sundqvist}, {Puls}, {Caballero-Nieves}, {Abdul-Masih},
  {Driessen}, {Garc{\'\i}a}, {Geen}, {Gr{\"a}fener}, {Hawcroft}, {Kaper},
  {Keszthelyi}, {Langer}, {Sana}, {Schneider}, {Shenar}, \& {Vink}}]{BdKB22}
{Brands}, S.~A., {de Koter}, A., {Bestenlehner}, J.~M., {et~al.} 2022, \aap,
  663, A36

\bibitem[{{Brott} {et~al.}(2011){Brott}, {de Mink}, {Cantiello}, {Langer}, {de
  Koter}, {Evans}, {Hunter}, {Trundle}, \& {Vink}}]{BdMC11}
{Brott}, I., {de Mink}, S.~E., {Cantiello}, M., {et~al.} 2011, \aap, 530, A115

\bibitem[{{Cantat-Gaudin} \& {Brandt}(2021)}]{CGB21}
{Cantat-Gaudin}, T. \& {Brandt}, T.~D. 2021, \aap, 649, A124

\bibitem[{{Ceverino} \& {Klypin}(2009)}]{CeK09}
{Ceverino}, D. \& {Klypin}, A. 2009, \apj, 695, 292

\bibitem[{{Chatterjee} \& {Tan}(2012)}]{CT12}
{Chatterjee}, S. \& {Tan}, J.~C. 2012, \apj, 754, 152

\bibitem[{{Conroy} \& {Kratter}(2012)}]{CoK12}
{Conroy}, C. \& {Kratter}, K.~M. 2012, \apj, 755, 123

\bibitem[{{Cumming}(2004)}]{Cum04}
{Cumming}, A. 2004, \mnras, 354, 1165

\bibitem[{{de Mink} {et~al.}(2013){de Mink}, {Langer}, {Izzard}, {Sana}, \& {de
  Koter}}]{dMLI13}
{de Mink}, S.~E., {Langer}, N., {Izzard}, R.~G., {Sana}, H., \& {de Koter}, A.
  2013, \apj, 764, 166

\bibitem[{{Dorigo Jones} {et~al.}(2020){Dorigo Jones}, {Oey}, {Paggeot},
  {Castro}, \& {Moe}}]{JOP20}
{Dorigo Jones}, J., {Oey}, M.~S., {Paggeot}, K., {Castro}, N., \& {Moe}, M.
  2020, \apj, 903, 43

\bibitem[{{Eldridge} {et~al.}(2011){Eldridge}, {Langer}, \& {Tout}}]{ELT11}
{Eldridge}, J.~J., {Langer}, N., \& {Tout}, C.~A. 2011, \mnras, 414, 3501

\bibitem[{{Evans} {et~al.}(2011){Evans}, {Taylor}, {H{\'e}nault-Brunet},
  {Sana}, {de Koter}, {Sim{\'o}n-D{\'{\i}}az}, {Carraro}, {Bagnoli}, {Bastian},
  {Bestenlehner}, {Bonanos}, {Bressert}, {Brott}, {Campbell}, {Cantiello},
  {Clark}, {Costa}, {Crowther}, {de Mink}, {Doran}, {Dufton}, {Dunstall},
  {Friedrich}, {Garcia}, {Gieles}, {Gr{\"a}fener}, {Herrero}, {Howarth},
  {Izzard}, {Langer}, {Lennon}, {Ma{\'{\i}}z Apell{\'a}niz}, {Markova},
  {Najarro}, {Puls}, {Ramirez}, {Sab{\'{\i}}n-Sanjuli{\'a}n}, {Smartt},
  {Stroud}, {van Loon}, {Vink}, \& {Walborn}}]{ETHB11}
{Evans}, C.~J., {Taylor}, W.~D., {H{\'e}nault-Brunet}, V., {et~al.} 2011, \aap,
  530, A108

\bibitem[{{Evans} {et~al.}(2010){Evans}, {Walborn}, {Crowther},
  {H{\'e}nault-Brunet}, {Massa}, {Taylor}, {Howarth}, {Sana}, {Lennon}, \& {van
  Loon}}]{EWC10}
{Evans}, C.~J., {Walborn}, N.~R., {Crowther}, P.~A., {et~al.} 2010, \apjl, 715,
  L74

\bibitem[{{Evans} {et~al.}(2020){Evans}, {Renzo}, \& {Rossi}}]{ERR20}
{Evans}, F.~A., {Renzo}, M., \& {Rossi}, E.~M. 2020, \mnras, 497, 5344

\bibitem[{{Fujii} \& {Portegies Zwart}(2011)}]{FPZ11}
{Fujii}, M.~S. \& {Portegies Zwart}, S. 2011, Science, 334, 1380

\bibitem[{{Gaia Collaboration} {et~al.}(2022){Gaia Collaboration}, {Vallenari},
  {Brown}, {Prusti}, {de Bruijne}, {Arenou}, {Babusiaux}, {Biermann},
  {Creevey}, {Ducourant}, {Evans}, {Eyer}, {Guerra}, {Hutton}, {Jordi},
  {Klioner}, {Lammers}, {Lindegren}, {Luri}, {Mignard}, {Panem}, {Pourbaix},
  {Randich}, {Sartoretti}, {Soubiran}, {Tanga}, {Walton}, {Bailer-Jones},
  {Bastian}, {Drimmel}, {Jansen}, {Katz}, {Lattanzi}, {van Leeuwen}, {Bakker},
  {Cacciari}, {Casta{\~n}eda}, {De Angeli}, {Fabricius}, {Fouesneau},
  {Fr{\'e}mat}, {Galluccio}, {Guerrier}, {Heiter}, {Masana}, {Messineo},
  {Mowlavi}, {Nicolas}, {Nienartowicz}, {Pailler}, {Panuzzo}, {Riclet}, {Roux},
  {Seabroke}, {Sordo{\o}rcit}, {Th{\'e}venin}, {Gracia-Abril}, {Portell},
  {Teyssier}, {Altmann}, {Andrae}, {Audard}, {Bellas-Velidis}, {Benson},
  {Berthier}, {Blomme}, {Burgess}, {Busonero}, {Busso}, {C{\'a}novas}, {Carry},
  {Cellino}, {Cheek}, {Clementini}, {Damerdji}, {Davidson}, {de Teodoro},
  {Nu{\~n}ez Campos}, {Delchambre}, {Dell'Oro}, {Esquej},
  {Fern{\'a}ndez-Hern{\'a}ndez}, {Fraile}, {Garabato}, {Garc{\'\i}a-Lario},
  {Gosset}, {Haigron}, {Halbwachs}, {Hambly}, {Harrison}, {Hern{\'a}ndez},
  {Hestroffer}, {Hodgkin}, {Holl}, {Jan{\ss}en}, {Jevardat de Fombelle},
  {Jordan}, {Krone-Martins}, {Lanzafame}, {L{\"o}ffler}, {Marchal}, {Marrese},
  {Moitinho}, {Muinonen}, {Osborne}, {Pancino}, {Pauwels}, {Recio-Blanco},
  {Reyl{\'e}}, {Riello}, {Rimoldini}, {Roegiers}, {Rybizki}, {Sarro}, {Siopis},
  {Smith}, {Sozzetti}, {Utrilla}, {van Leeuwen}, {Abbas}, {{\'A}brah{\'a}m},
  {Abreu Aramburu}, {Aerts}, {Aguado}, {Ajaj}, {Aldea-Montero}, {Altavilla},
  {{\'A}lvarez}, {Alves}, {Anders}, {Anderson}, {Anglada Varela}, {Antoja},
  {Baines}, {Baker}, {Balaguer-N{\'u}{\~n}ez}, {Balbinot}, {Balog}, {Barache},
  {Barbato}, {Barros}, {Barstow}, {Bartolom{\'e}}, {Bassilana}, {Bauchet},
  {Becciani}, {Bellazzini}, {Berihuete}, {Bernet}, {Bertone}, {Bianchi},
  {Binnenfeld}, {Blanco-Cuaresma}, {Blazere}, {Boch}, {Bombrun}, {Bossini},
  {Bouquillon}, {Bragaglia}, {Bramante}, {Breedt}, {Bressan}, {Brouillet},
  {Brugaletta}, {Bucciarelli}, {Burlacu}, {Butkevich}, {Buzzi}, {Caffau},
  {Cancelliere}, {Cantat-Gaudin}, {Carballo}, {Carlucci}, {Carnerero},
  {Carrasco}, {Casamiquela}, {Castellani}, {Castro-Ginard}, {Chaoul},
  {Charlot}, {Chemin}, {Chiaramida}, {Chiavassa}, {Chornay}, {Comoretto},
  {Contursi}, {Cooper}, {Cornez}, {Cowell}, {Crifo}, {Cropper}, {Crosta},
  {Crowley}, {Dafonte}, {Dapergolas}, {David}, {David}, {de Laverny}, {De
  Luise}, {De March}, {De Ridder}, {de Souza}, {de Torres}, {del Peloso}, {del
  Pozo}, {Delbo}, {Delgado}, {Delisle}, {Demouchy}, {Dharmawardena}, {Di
  Matteo}, {Diakite}, {Diener}, {Distefano}, {Dolding}, {Edvardsson}, {Enke},
  {Fabre}, {Fabrizio}, {Faigler}, {Fedorets}, {Fernique}, {Fienga}, {Figueras},
  {Fournier}, {Fouron}, {Fragkoudi}, {Gai}, {Garcia-Gutierrez},
  {Garcia-Reinaldos}, {Garc{\'\i}a-Torres}, {Garofalo}, {Gavel}, {Gavras},
  {Gerlach}, {Geyer}, {Giacobbe}, {Gilmore}, {Girona}, {Giuffrida}, {Gomel},
  {Gomez}, {Gonz{\'a}lez-N{\'u}{\~n}ez}, {Gonz{\'a}lez-Santamar{\'\i}a},
  {Gonz{\'a}lez-Vidal}, {Granvik}, {Guillout}, {Guiraud},
  {Guti{\'e}rrez-S{\'a}nchez}, {Guy}, {Hatzidimitriou}, {Hauser}, {Haywood},
  {Helmer}, {Helmi}, {Sarmiento}, {Hidalgo}, {Hilger}, {H{\l}adczuk}, {Hobbs},
  {Holland}, {Huckle}, {Jardine}, {Jasniewicz}, {Jean-Antoine Piccolo},
  {Jim{\'e}nez-Arranz}, {Jorissen}, {Juaristi Campillo}, {Julbe}, {Karbevska},
  {Kervella}, {Khanna}, {Kontizas}, {Kordopatis}, {Korn}, {K{\'o}sp{\'a}l},
  {Kostrzewa-Rutkowska}, {Kruszy{\'n}ska}, {Kun}, {Laizeau}, {Lambert},
  {Lanza}, {Lasne}, {Le Campion}, {Lebreton}, {Lebzelter}, {Leccia}, {Leclerc},
  {Lecoeur-Taibi}, {Liao}, {Licata}, {Lindstr{\o}m}, {Lister}, {Livanou},
  {Lobel}, {Lorca}, {Loup}, {Madrero Pardo}, {Magdaleno Romeo}, {Managau},
  {Mann}, {Manteiga}, {Marchant}, {Marconi}, {Marcos}, {Marcos Santos},
  {Mar{\'\i}n Pina}, {Marinoni}, {Marocco}, {Marshall}, {Polo},
  {Mart{\'\i}n-Fleitas}, {Marton}, {Mary}, {Masip}, {Massari},
  {Mastrobuono-Battisti}, {Mazeh}, {McMillan}, {Messina}, {Michalik}, {Millar},
  {Mints}, {Molina}, {Molinaro}, {Moln{\'a}r}, {Monari}, {Mongui{\'o}},
  {Montegriffo}, {Montero}, {Mor}, {Mora}, {Morbidelli}, {Morel}, {Morris},
  {Muraveva}, {Murphy}, {Musella}, {Nagy}, {Noval}, {Oca{\~n}a}, {Ogden},
  {Ordenovic}, {Osinde}, {Pagani}, {Pagano}, {Palaversa}, {Palicio},
  {Pallas-Quintela}, {Panahi}, {Payne-Wardenaar}, {Pe{\~n}alosa Esteller},
  {Penttil{\"a}}, {Pichon}, {Piersimoni}, {Pineau}, {Plachy}, {Plum}, {Poggio},
  {Pr{\v{s}}a}, {Pulone}, {Racero}, {Ragaini}, {Rainer}, {Raiteri}, {Rambaux},
  {Ramos}, {Ramos-Lerate}, {Re Fiorentin}, {Regibo}, {Richards}, {Rios Diaz},
  {Ripepi}, {Riva}, {Rix}, {Rixon}, {Robichon}, {Robin}, {Robin}, {Roelens},
  {Rogues}, {Rohrbasser}, {Romero-G{\'o}mez}, {Rowell}, {Royer}, {Ruz Mieres},
  {Rybicki}, {Sadowski}, {S{\'a}ez N{\'u}{\~n}ez}, {Sagrist{\`a} Sell{\'e}s},
  {Sahlmann}, {Salguero}, {Samaras}, {Sanchez Gimenez}, {Sanna},
  {Santove{\~n}a}, {Sarasso}, {Schultheis}, {Sciacca}, {Segol}, {Segovia},
  {S{\'e}gransan}, {Semeux}, {Shahaf}, {Siddiqui}, {Siebert}, {Siltala},
  {Silvelo}, {Slezak}, {Slezak}, {Smart}, {Snaith}, {Solano}, {Solitro},
  {Souami}, {Souchay}, {Spagna}, {Spina}, {Spoto}, {Steele},
  {Steidelm{\"u}ller}, {Stephenson}, {S{\"u}veges}, {Surdej}, {Szabados},
  {Szegedi-Elek}, {Taris}, {Taylo}, {Teixeira}, {Tolomei}, {Tonello}, {Torra},
  {Torra}, {Torralba Elipe}, {Trabucchi}, {Tsounis}, {Turon}, {Ulla}, {Unger},
  {Vaillant}, {van Dillen}, {van Reeven}, {Vanel}, {Vecchiato}, {Viala},
  {Vicente}, {Voutsinas}, {Weiler}, {Wevers}, {Wyrzykowski}, {Yoldas}, {Yvard},
  {Zhao}, {Zorec}, {Zucker}, \& {Zwitter}}]{GaiaDR3}
{Gaia Collaboration}, {Vallenari}, A., {Brown}, A.~G.~A., {et~al.} 2022, arXiv
  e-prints, arXiv:2208.00211

\bibitem[{{Gies} \& {Bolton}(1986)}]{GiB86}
{Gies}, D.~R. \& {Bolton}, C.~T. 1986, \apjs, 61, 419

\bibitem[{{H{\'e}nault-Brunet} {et~al.}(2012){H{\'e}nault-Brunet}, {Evans},
  {Sana}, {Gieles}, {Bastian}, {Ma{\'\i}z Apell{\'a}niz}, {Markova}, {Taylor},
  {Bressert}, {Crowther}, \& {van Loon}}]{HBES12}
{H{\'e}nault-Brunet}, V., {Evans}, C.~J., {Sana}, H., {et~al.} 2012, \aap, 546,
  A73

\bibitem[{{Hoogerwerf} {et~al.}(2000){Hoogerwerf}, {de Bruijne}, \& {de
  Zeeuw}}]{HdBdZ00}
{Hoogerwerf}, R., {de Bruijne}, J.~H.~J., \& {de Zeeuw}, P.~T. 2000, \apjl,
  544, L133

\bibitem[{{Hoogerwerf} {et~al.}(2001){Hoogerwerf}, {de Bruijne}, \& {de
  Zeeuw}}]{HdBdZ01}
{Hoogerwerf}, R., {de Bruijne}, J.~H.~J., \& {de Zeeuw}, P.~T. 2001, \aap, 365,
  49

\bibitem[{{Jilinski} {et~al.}(2010){Jilinski}, {Ortega}, {Drake}, \& {de la
  Reza}}]{JOD10}
{Jilinski}, E., {Ortega}, V.~G., {Drake}, N.~A., \& {de la Reza}, R. 2010,
  \apj, 721, 469

\bibitem[{{Lebouteiller} {et~al.}(2019){Lebouteiller}, {Cormier}, {Madden},
  {Galametz}, {Hony}, {Galliano}, {Chevance}, {Lee}, {Braine}, {Polles},
  {Reque{\~n}a-Torres}, {Indebetouw}, {Hughes}, \& {Abel}}]{LCM19}
{Lebouteiller}, V., {Cormier}, D., {Madden}, S.~C., {et~al.} 2019, \aap, 632,
  A106

\bibitem[{{Lennon} {et~al.}(2018){Lennon}, {Evans}, {van der Marel},
  {Anderson}, {Platais}, {Herrero}, {de Mink}, {Sana}, {Sabbi}, {Bedin},
  {Crowther}, {Langer}, {Ramos Lerate}, {del Pino}, {Renzo},
  {Sim{\'o}n-D{\'\i}az}, \& {Schneider}}]{LEvdM18}
{Lennon}, D.~J., {Evans}, C.~J., {van der Marel}, R.~P., {et~al.} 2018, \aap,
  619, A78

\bibitem[{{Mahy} {et~al.}(2020{\natexlab{a}}){Mahy}, {Almeida}, {Sana},
  {Clark}, {de Koter}, {de Mink}, {Evans}, {Grin}, {Langer}, {Moffat},
  {Schneider}, {Shenar}, \& {Tramper}}]{MAS20}
{Mahy}, L., {Almeida}, L.~A., {Sana}, H., {et~al.} 2020{\natexlab{a}}, \aap,
  634, A119

\bibitem[{{Mahy} {et~al.}(2020{\natexlab{b}}){Mahy}, {Sana}, {Abdul-Masih},
  {Almeida}, {Langer}, {Shenar}, {de Koter}, {de Mink}, {de Wit}, {Grin},
  {Evans}, {Moffat}, {Schneider}, {Barb{\'a}}, {Clark}, {Crowther},
  {Gr{\"a}fener}, {Lennon}, {Tramper}, \& {Vink}}]{MSA20}
{Mahy}, L., {Sana}, H., {Abdul-Masih}, M., {et~al.} 2020{\natexlab{b}}, \aap,
  634, A118

\bibitem[{{Mahy} {et~al.}(2022){Mahy}, {Sana}, {Shenar}, {Sen}, {Langer},
  {Marchant}, {Abdul-Masih}, {Banyard}, {Bodensteiner}, {Bowman}, {Dsilva},
  {Fabry}, {Hawcroft}, {Janssens}, {Van Reeth}, \& {Eldridge}}]{MSS22}
{Mahy}, L., {Sana}, H., {Shenar}, T., {et~al.} 2022, \aap, 664, A159

\bibitem[{{Ma{\'\i}z Apell{\'a}niz}(2022)}]{JMA22}
{Ma{\'\i}z Apell{\'a}niz}, J. 2022, \aap, 657, A130

\bibitem[{{Ma{\'{\i}}z Apell{\'a}niz} {et~al.}(2014){Ma{\'{\i}}z
  Apell{\'a}niz}, {Evans}, {Barb{\'a}}, {Gr{\"a}fener}, {Bestenlehner},
  {Crowther}, {Garc{\'{\i}}a}, {Herrero}, {Sana}, {Sim{\'o}n-D{\'{\i}}az},
  {Taylor}, {van Loon}, {Vink}, \& {Walborn}}]{MAEB14}
{Ma{\'{\i}}z Apell{\'a}niz}, J., {Evans}, C.~J., {Barb{\'a}}, R.~H., {et~al.}
  2014, \aap, 564, A63

\bibitem[{{Ma{\'\i}z Apell{\'a}niz} {et~al.}(2018){Ma{\'\i}z Apell{\'a}niz},
  {Pantaleoni Gonz{\'a}lez}, {Barb{\'a}}, {Sim{\'o}n-D{\'\i}az}, {Negueruela},
  {Lennon}, {Sota}, \& {Trigueros P{\'a}ez}}]{MAPGB18}
{Ma{\'\i}z Apell{\'a}niz}, J., {Pantaleoni Gonz{\'a}lez}, M., {Barb{\'a}},
  R.~H., {et~al.} 2018, \aap, 616, A149

\bibitem[{{Pasquini} {et~al.}(2002){Pasquini}, {Avila}, {Blecha}, {Cacciari},
  {Cayatte}, {Colless}, {Damiani}, {de Propris}, {Dekker}, {di Marcantonio},
  {Farrell}, {Gillingham}, {Guinouard}, {Hammer}, {Kaufer}, {Hill}, {Marteaud},
  {Modigliani}, {Mulas}, {North}, {Popovic}, {Rossetti}, {Royer}, {Santin},
  {Schmutzer}, {Simond}, {Vola}, {Waller}, \& {Zoccali}}]{PAB02}
{Pasquini}, L., {Avila}, G., {Blecha}, A., {et~al.} 2002, The Messenger, 110, 1

\bibitem[{{Platais} {et~al.}(2018){Platais}, {Lennon}, {van der Marel},
  {Bellini}, {Sabbi}, {Watkins}, {Sohn}, {Walborn}, {Bedin}, {Evans}, {de
  Mink}, {Sana}, {Herrero}, {Langer}, \& {Crowther}}]{PLvdM18}
{Platais}, I., {Lennon}, D.~J., {van der Marel}, R.~P., {et~al.} 2018, \aj,
  156, 98

\bibitem[{{Ram{\'{\i}}rez-Agudelo} {et~al.}(2013){Ram{\'{\i}}rez-Agudelo},
  {Sim{\'o}n-D{\'{\i}}az}, {Sana}, {de Koter}, {Sab{\'{\i}}n-Sanjul{\'{\i}}an},
  {de Mink}, {Dufton}, {Gr{\"a}fener}, {Evans}, {Herrero}, {Langer}, {Lennon},
  {Ma{\'{\i}}z Apell{\'a}niz}, {Markova}, {Najarro}, {Puls}, {Taylor}, \&
  {Vink}}]{RASDS13}
{Ram{\'{\i}}rez-Agudelo}, O.~H., {Sim{\'o}n-D{\'{\i}}az}, S., {Sana}, H.,
  {et~al.} 2013, \aap, 560, A29

\bibitem[{{Renzo} {et~al.}(2019){Renzo}, {Zapartas}, {de Mink}, {G{\"o}tberg},
  {Justham}, {Farmer}, {Izzard}, {Toonen}, \& {Sana}}]{RZdM19}
{Renzo}, M., {Zapartas}, E., {de Mink}, S.~E., {et~al.} 2019, \aap, 624, A66

\bibitem[{{Sabbi} {et~al.}(2012){Sabbi}, {Lennon}, {Gieles}, {de Mink},
  {Walborn}, {Anderson}, {Bellini}, {Panagia}, {van der Marel}, \& {Ma{\'{\i}}z
  Apell{\'a}niz}}]{SLG12}
{Sabbi}, E., {Lennon}, D.~J., {Gieles}, M., {et~al.} 2012, \apjl, 754, L37

\bibitem[{{Sab{\'{\i}}n-Sanjuli{\'a}n}
  {et~al.}(2014){Sab{\'{\i}}n-Sanjuli{\'a}n}, {Sim{\'o}n-D{\'{\i}}az},
  {Herrero}, {Walborn}, {Puls}, {Ma{\'{\i}}z Apell{\'a}niz}, {Evans}, {Brott},
  {de Koter}, {Garcia}, {Markova}, {Najarro}, {Ram{\'{\i}}rez-Agudelo}, {Sana},
  {Taylor}, \& {Vink}}]{SSSDH14}
{Sab{\'{\i}}n-Sanjuli{\'a}n}, C., {Sim{\'o}n-D{\'{\i}}az}, S., {Herrero}, A.,
  {et~al.} 2014, \aap, 564, A39

\bibitem[{{Sana} {et~al.}(2013){Sana}, {de Koter}, {de Mink}, {Dunstall},
  {Evans}, {H{\'e}nault-Brunet}, {Ma{\'{\i}}z Apell{\'a}niz},
  {Ram{\'{\i}}rez-Agudelo}, {Taylor}, {Walborn}, {Clark}, {Crowther},
  {Herrero}, {Gieles}, {Langer}, {Lennon}, \& {Vink}}]{SdKdM13}
{Sana}, H., {de Koter}, A., {de Mink}, S.~E., {et~al.} 2013, \aap, 550, A107

\bibitem[{{Schneider} {et~al.}(2020){Schneider}, {Ohlmann}, {Podsiadlowski},
  {R{\"o}pke}, {Balbus}, \& {Pakmor}}]{SOP20}
{Schneider}, F.~R.~N., {Ohlmann}, S.~T., {Podsiadlowski}, P., {et~al.} 2020,
  \mnras, 495, 2796

\bibitem[{{Schneider} {et~al.}(2019){Schneider}, {Ohlmann}, {Podsiadlowski},
  {R{\"o}pke}, {Balbus}, {Pakmor}, \& {Springel}}]{SOP19}
{Schneider}, F. R.~N., {Ohlmann}, S.~T., {Podsiadlowski}, P., {et~al.} 2019,
  \nat, 574, 211

\bibitem[{{Schneider} {et~al.}(2018){Schneider}, {Sana}, {Evans},
  {Bestenlehner}, {Castro}, {Fossati}, {Gr{\"a}fener}, {Langer},
  {Ram{\'{\i}}rez-Agudelo}, {Sab{\'{\i}}n-Sanjuli{\'a}n},
  {Sim{\'o}n-D{\'{\i}}az}, {Tramper}, {Crowther}, {de Koter}, {de Mink},
  {Dufton}, {Garcia}, {Gieles}, {H{\'e}nault-Brunet}, {Herrero}, {Izzard},
  {Kalari}, {Lennon}, {Ma{\'{\i}}z Apell{\'a}niz}, {Markova}, {Najarro},
  {Podsiadlowski}, {Puls}, {Taylor}, {van Loon}, {Vink}, \& {Norman}}]{SSE18}
{Schneider}, F.~R.~N., {Sana}, H., {Evans}, C.~J., {et~al.} 2018, Science, 359,
  69

\bibitem[{{Sen} {et~al.}(2022){Sen}, {Langer}, {Marchant}, {Menon}, {de Mink},
  {Schootemeijer}, {Sch{\"u}rmann}, {Mahy}, {Hastings}, {Nathaniel}, {Sana},
  {Wang}, \& {Xu}}]{SLM2022}
{Sen}, K., {Langer}, N., {Marchant}, P., {et~al.} 2022, \aap, 659, A98

\bibitem[{{Shenar} {et~al.}(2022{\natexlab{a}}){Shenar}, {Sana}, {Mahy},
  {El-Badry}, {Marchant}, {Langer}, {Hawcroft}, {Fabry}, {Sen}, {Almeida},
  {Abdul-Masih}, {Bodensteiner}, {Crowther}, {Gieles}, {Gromadzki},
  {H{\'e}nault-Brunet}, {Herrero}, {Koter}, {Iwanek}, {Koz{\l}owski}, {Lennon},
  {Apell{\'a}niz}, {Mr{\'o}z}, {Moffat}, {Picco}, {Pietrukowicz}, {Poleski},
  {Rybicki}, {Schneider}, {Skowron}, {Skowron}, {Soszy{\'n}ski},
  {Szyma{\'n}ski}, {Toonen}, {Udalski}, {Ulaczyk}, {Vink}, \&
  {Wrona}}]{SSM2022a}
{Shenar}, T., {Sana}, H., {Mahy}, L., {et~al.} 2022{\natexlab{a}}, Nature
  Astronomy

\bibitem[{{Shenar} {et~al.}(2022{\natexlab{b}}){Shenar}, {Sana}, {Mahy},
  {Ma{\'\i}z Apell{\'a}niz}, {Crowther}, {Gromadzki}, {Herrero}, {Langer},
  {Marchant}, {Schneider}, {Sen}, {Soszy{\'n}ski}, \& {Toonen}}]{SSM2022b}
{Shenar}, T., {Sana}, H., {Mahy}, L., {et~al.} 2022{\natexlab{b}}, \aap, 665,
  A148

\bibitem[{{Sim{\'o}n-D{\'\i}az} {et~al.}(2020){Sim{\'o}n-D{\'\i}az},
  {Britvaskiy}, {Castro}, \& {Holgado}}]{SSBC20}
{Sim{\'o}n-D{\'\i}az}, S., {Britvaskiy}, N., {Castro}, N., \& {Holgado}, G.
  2020, in XIV.0 Scientific Meeting (virtual) of the Spanish Astronomical
  Society, 186

\bibitem[{{Stone}(1991)}]{Sto91}
{Stone}, R.~C. 1991, \aj, 102, 333

\bibitem[{{Walborn} {et~al.}(2014){Walborn}, {Sana}, {Sim{\'o}n-D{\'{\i}}az},
  {Ma{\'{\i}}z Apell{\'a}niz}, {Taylor}, {Evans}, {Markova}, {Lennon}, \& {de
  Koter}}]{WSSD14}
{Walborn}, N.~R., {Sana}, H., {Sim{\'o}n-D{\'{\i}}az}, S., {et~al.} 2014, \aap,
  564, A40

\bibitem[{{Zwicky}(1957)}]{Zwi57}
{Zwicky}, F. 1957, {Morphological astronomy}

\end{thebibliography}

\begin{appendix}

\section{Gaia DR3 proper motions} \label{sect: gaia}

\begin{figure*}
\centering
\includegraphics[width=\textwidth]{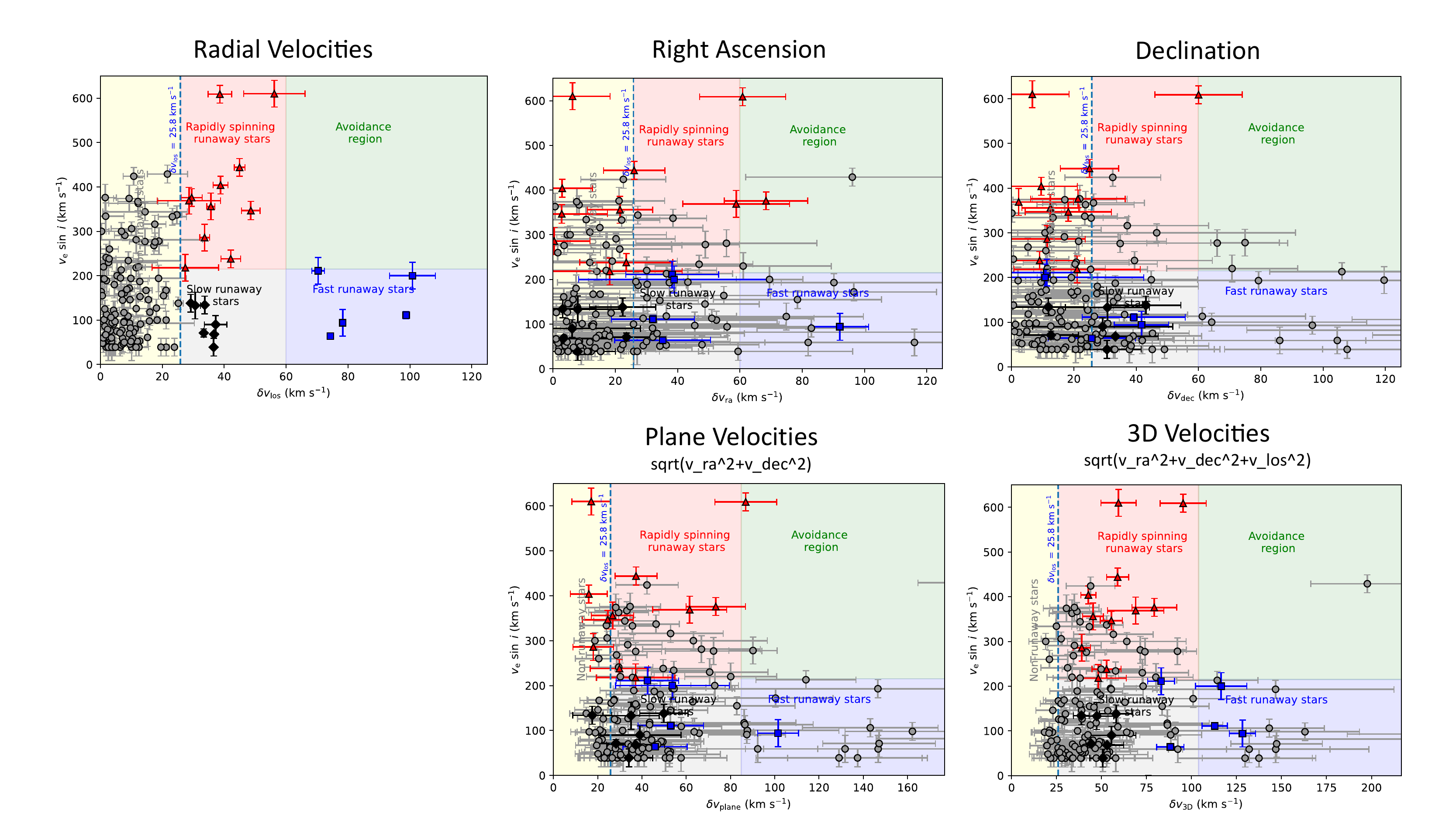}
\caption{Diagram of the projected rotation velocities ($v_\mathrm{e} \sin i$) versus peculiar 1D space velocities (along the line of sight, RA and DEC on the top row), as well as plane (2D) and 3D peculiar velocities (bottom row) for the full VFTS single O star sample.  We exclude 32 objects with GAIA RUWE above 1.4, as this potentially indicate a poor fit or an unresolved system. The boundary of the RW desert obtained via line-of-sight velocities ($v_\mathrm{e} \sin i > 210$~\kms\ and $\delta v_\mathrm{los}> 60$~\kms) have been corrected on the bottom row to account for the 2D and 3D nature of the observational constraints using scaling factors of $\sqrt{2}$ and $\sqrt{3}$, respectively. The one object in the avoidance region in 3D-velocities panel has a $\delta v_{dec} \sim 175$~\kms\ and  $\delta v_{plane} \sim 200$~\kms, i.e.\ so large that they do not appear on the respective panels.}
\label{fig:figS7}
\end{figure*}


We retrieved the \gaia\ measurements of the 180 VFTS presumably single O stars   from the ARI'Heidelberg Data Release 3 archive \citep{GaiaDR3} and applied the corrections recommended by \citet{CGB21} and \citet{JMA22}. We rejected 32 stars with Gaia Renormalised Unit Weight Error (RUWE)  fit quality metric above 1.4, 

The median measurement uncertainty for the \gaia\ corrected proper motions is 0.07~\masyr\ for our sample, equivalent to $\sim$16~\kms\ given the distance to 30~Dor. The proper motion errors are thus significantly larger than the errors on the line-of-sight velocities. Yet, they are small enough to offer a first challenge to the zone of avoidance identified in Fig~\ref{fig: RWdesert}. 
As for the line of sight velocities, we computed the star proper motion with respect to the central region by subtracting the mean proper motion estimated from the median of 70 stars in a 5\arcsec\ radius around R136, with good RUWE and parallaxes estimate between 0.017 and 0.025~mas: $pmra_\mathrm{R136}=1.7003030$~\masyr\ and $pmdec_\mathrm{R136}=0.68414241$~\masyr. This value is slightly different ($\sim 10$~\kms) than that proposed by \citep{LEvdM18} $pmra_\mathrm{R136}=1.73863$~\masyr\ and $pmdec_\mathrm{R136}=0.701245$~\masyr) but is conveniently between the value of 
\citep{LEvdM18} and that obtained from the median of DR3 propper motion of the VFTS single O star sample. Given the likely velocity structure of different regions of the Tarantula nebula (see Table~\ref{tab: cluster}, a better precision would require an in-depth investigation that is beyond the present goals. 

Figure~\ref{fig:figS7} (upper row) shows that the  runaway desert remains under-populated in
 the rotational velocity $vs.$  right-ascension and declination space velocities. 
We also computed the space velocities in the tangential plane and the 3D space velocities. 
Results are displayed in Figure~\ref{fig:figS7} (bottom row). Unfortunately, the quality of \gaia\ data for the O-type stars is rather inhomogeneous and likely affected by crowding issue for a significant number of objects. Object with genuinely large plane-of-sky velocities ($ > 100$~\kms) are challenging  to disentangle from objects suffering from crowdedness and other measurement biases \citep{LEvdM18}. The completeness level of {\it good} measurements is even harder to asses so that  reliable assessment of observational biases based on GAIA data in 30~Dor will have to wait further investigations. Yet, the region of avoidance remains underpopulated but for one object (VFTS~615). If the large plane velocity of VFTS~615 ($\sim 200$~\kms) is confirmed, this would be the fastest RW object identified in 30~Dor. The reality of this latter findings is far from obvious. Indeed, a large proper motion study comparing HST and Gaia proper motion do not find a large number of additional rapidly-moving massive runaways \citep{PLvdM18}, confirming once more that our present results are not strongly biased against such objects. 

\end{appendix}

\end{document}